\documentclass[fleqn,usenatbib]{mnras}
\usepackage{graphicx}
\usepackage{url}
\usepackage{longtable}
\usepackage{float}
\usepackage{subfigure}
\usepackage{newtxtext,newtxmath}
\usepackage[T1]{fontenc}
\usepackage{amsmath}
\usepackage{soul}
\usepackage{lscape}

\DeclareRobustCommand{\VAN}[3]{#2} \let\VANthebibliography\thebibliography
\def\thebibliography{\DeclareRobustCommand{\VAN}[3]{##3} \VANthebibliography}

\title[3D velocities of YSO-MC associations]{Kinematics of the interstellar medium using Gaia: A catalogue of 102 YSO-MC associations within 3.5 kpc from the Sun with 3D velocities}

\author[Zhou, Li \& Chen]{
Ji-Xuan Zhou,$^{1}$,$^{2}$
Guang-Xing Li,$^{1}$\thanks{gxli@ynu.edu.cn} 
Bing-Qiu Chen$^{1}$\thanks{bchen@ynu.edu.cn} \\
$^{1}$ South-Western Institute for Astronomy Research, Yunnan University, Chenggong District, Kunming 650091, P.\,R. China \\
$^{2}$ School of Physics and Astronomy, Cardiff University, Queen's Buildings, The Parade, Cardiff, CF24 3AA, UK}

\begin{document}
\label{firstpage}

\maketitle

\begin{abstract}
Kinematic information is crucial for understanding the evolution of complex
systems, such as interstellar gas. Obtaining full 3D kinematic information is a
crucial final step for modeling and interpretation. Molecular clouds are
nurseries where stars are born. Stars at a very early stage, like young stellar
objects (YSOs), inherit the spatial and kinematic structure of the gas patches
they originate from. In this paper, we combine measurements of radial velocities
towards the gas and the kinematic information of YSOs from Gaia DR3 to derive 3D
velocities of a sample of YSO (Young Stellar Object)-MC (Molecular Cloud)
complexes at d$\lesssim$3.5kpc from the Sun. We find that the molecular
interstellar medium traced by the YSO-MC complexes generally follows Galactic
rotation, with an additional peculiar velocity of 8.6 km s$^{-1}$. The random
motion of these complexes in the Galactic XY plane is more energetic than motion
along the Z direction. A catalogue containing the 3D velocities of the YSO-MC
complexes at different reference frames is available, and the distances and 3D
velocities of well-known molecular clouds are presented. Our results set the
foundation for exploring the interplay between the Galaxy, the molecular ISM,
and star formation. Data available at \url{https://doi.org/10.5281/zenodo.16364877}.

\end{abstract} 

\begin{keywords} % commenting results in missing thanks notes!
    Galaxies: ISM  -- ISM: structure -- ISM: clouds  -- Stars: formation  --  ISM: kinematics and dynamics
\end{keywords}
% \noindent {\bf Key words:} 

\section{Introduction}\label{intro}

Molecular clouds are the birthplace of stars. They contain most cold, molecular gas in our Galaxy \citep{2001ApJ...547..792D}. These clouds are not isolated objects in galaxies but participate in the material cycle in galaxies. The formation of molecular clouds is a direct consequence of the evolution of the Galactic disc \citep[e.g.][]{2001MNRAS.327..663P}. During the cloud lifetime, Galactic-scale processes constantly affect cloud evolution by setting the boundary conditions where mass and kinematic energy are injected. The motion of molecular clouds is affected by many physical processes like Galactic shear \citep{2013A&A...559A..34L,2014ApJ...797...53G,2014A&A...568A..73R,2015MNRAS.447.3390D,2015MNRAS.450.4043W,2020ApJ...897...89L,2022MNRAS.516L..35L}, stellar feedback \citep{2009ApJ...695..292C}, and spiral density waves \citep{1969ApJ...155..721L,2014PASA...31...35D}. Besides a differential rotation caused by gravity from the dark matter, stars and molecular clouds still contain a significant amount of random motions.

Kinematic information is crucial to understanding the interplay between the clouds and the Galaxy, and full, 3D kinematic measurements of molecular clouds are crucial for understanding their evolution in the Galaxy. Earlier studies have identified many molecular clouds from CO emission maps, where properties of clouds like size, mass, radial velocity, velocity dispersions, and CO luminosities can be measured \citep{1975ApJ...199L.105S,1987ApJ...319..730S,2010ApJ...723..492R,2020MNRAS.493..351C}. The kinematic measurements derived using velocity-resolved CO measurements or other molecular lines are limited to 1D velocity along the radial direction.
Young stellar objects (YSOs) are stars in the earliest stages of evolution and are often used as tracers of molecular clouds. Based on their near- to mid-infrared spectral index, YSOs are classified into Class 0, I, II, and III \citep{Lada1987}. Through this sequence, young stars are classified from being deeply embedded in protostellar cores at the Class 0 stage to being surrounded by the envelope (Class I), circumstellar discs (Class II), and then with SEDs of stellar photospheres at the Class III stage \citep{Dunham2014}. Being embedded in molecular clouds or bearing discs or envelopes, YSOs exhibit infrared excesses, which are widely used to identify them. In addition, other observational features, e.g., outflows and jets driven by the accretion process \citep{Andre1993, Arce2007, Frank2014}, and X-ray emission generated from the interaction between the disc and the magnetic field \citep{Neuh1997, Feigelson1999}, are also used to characterize YSOs.

With the increasing number of young stellar objects (YSOs) identified since the launch of Spitzer, Herschel, and the Wide-field Infrared Survey Explorer (WISE) \citep{Evans2003, Evans2009, Andre2010, Fischer2016}, studies of their spatial distribution and kinematics have gained significant momentum \citep{Dunham2014}. Before molecular clouds are dispersed or expelled, YSOs remain spatially associated with them, sharing similar locations. Precise parallax measurements of YSOs have enabled accurate determinations of distances and positions for many molecular clouds \citep{Dzib2018, Zhang2023, Marton2023}. YSOs are also thought to inherit the velocities of their parent molecular clouds. Earlier studies comparing CO gas and YSO radial velocities in nearby clouds confirmed their dynamical linkage \citep{Covey2006}. Using high-resolution optical spectra from young stars in the Orion B molecular cloud, \citet{Kounkel2017} found that those in NGC 2068 and L1622 exhibit radial velocities similar to the molecular gas. Similarly, \citet{Josefa2021} reported close agreement between the mean radial velocities of CO gas and YSOs in the Orion molecular cloud. In more massive environments, such as the OB clusters in Cygnus, \citet{Quintana2022} demonstrated that the 3D velocity dispersion of expanding clusters follows Larson’s Law, indicating a connection to the turbulent dynamics of their parental clouds. \citet{Yang2025} found that radial velocity differences between CO gas and YSOs are typically less than 1.4 km s$^{-1}$ in five nearby star-forming regions (Orion A, Orion B, Perseus, Taurus, and $\lambda$ Orionis). Consequently, YSO kinematics are now widely utilized to investigate the motions and evolution of molecular clouds, including their expansion, structural evolution, and turbulence \citep{Dzib2018, Kounkel2018, Swiggum2025, Ha2021, 2022MNRAS.513..638Z}. In this work, we utilize YSO associations linked to molecular clouds, defined as YSO-MC (young stellar object-molecular cloud) complexes. These associations enable us to study the 3D velocity structure of molecular clouds, supplemented by 2D stellar velocity data from Gaia DR3.

The Gaia satellite \citep{2016A&A...595A...1G} provides high-precision astrometric measurements, enabling the determination of stellar distances, velocities, spectral types, and stellar parameters. The latest data release, Gaia DR3, includes observations of over 1.8 billion sources, offering improved parallaxes and proper motions, particularly for faint stars and a larger number of radial velocity measurements \citep{Gaia2023}. Full astrometric solutions are available for approximately 1.46 billion sources. Notably, the measurements of young stellar objects (YSOs) are especially valuable, as they can be used to trace the motion of the gas associated with them. In our previous work, we studied the kinematic properties of a sample of YSO associations and their gas counterparts \citep{2022MNRAS.513..638Z, 2024MNRAS.529.1091Z}. Based on their gas contents, we select a sample of YSO-MC complexes in which the gas and a group of YSOs are associated with each other. For these complexes, the full 3D velocities can be obtained by combining the radial velocities measured towards the gas with astrometry measurements towards the YSOs provided by the Gaia satellite. The scientific potential of this approach has been demonstrated in our previous papers \citep{2022MNRAS.516L..35L}. In this paper, we present a catalogue of YSO-MC complexes with full, 3D velocities in different reference frames located at d$\lesssim$3.5 kpc from the Sun. We present our calculations with full details and discuss the peculiar velocity of the YSO-MC complexes.

\section{Derivation of 3D Velocities}
\subsection{Assumptions on the Galactic dynamics}
Given the close relation between kinematics of YSOs and molecular clouds found in references in Sec. \ref{intro}, we made an assumption that the YSOs still keep the velocity inherited from their parent molecular clouds. We limited ourselves to certain kinds of YSO associations where YSOs have a similar spatial distribution with the gas, indicating they are young enough and associated with the gas. We took the mean proper motion in l\&b directions of the YSO associations as the velocity of the clouds. Then, the 2D velocities can be combined with the radial velocity of the gas to fully describe the cloud motion.

\subsection{YSO associations with Gaia kinematic measurements}
A sample containing 102 YSO associations is taken from our last work \citep{2022MNRAS.513..638Z}, where the YSOs used are Class I and Class II type \citep{2016MNRAS.458.3479M}. These YSO associations are extracted from their density map in l-b-log(d) space by using the \texttt{Dendrogram} method \citep{2008ApJ...679.1338R} \footnote{\url{http://www.dendrograms.org/}.}, defined as spatially and kinematically clustered YSO groups. The 102 associations contain large structures and denser substructures.

The YSO associations are classified into different types based on their relation with the molecular gas in their vicinity \citep{2024MNRAS.529.1091Z}. The 102 associations used in this work are all gas-rich, including the Type 1 direct association (where most of the YSOs are closely associated with a cloud), Type 2 close association (where part of the YSOs are associated with the clouds), and Type 3 bubble association (YSOs are located within bubble gas structures). As these are objects associated with the molecular gas, the radial velocity can be obtained reliably using gas tracers. In this paper, we call these objects YSO-MC complexes. To update the location and velocity information of the member YSOs, we cross-matched them with the Gaia DR3 catalogue, which provides more accurate photometric and kinematic measurements \citep{Gaia2023}. For each YSO-MC complex, the proper motion (pmra and pmdec) and 3D location (l, b, d) are obtained from Gaia DR3, and the proper motions in right ascension (RA) and declination (Dec) are later transformed into pm${l}$ and pm${b}$ in Galactic coordinates using \texttt{Galpy} \citep{2015ApJS..216...29B}.

\subsection{Radial velocity measurements from CO}
The radial velocity towards each cloud is derived from \textsuperscript{12}CO emission data \citep{2001ApJ...547..792D}. This composite CO dataset towards the Milky Way was assembled from 7 individual surveys. It has an angular resolution ranging from $9'$ to $18'$, with a velocity resolution of 1.3 km s$^{-1}$. The matching between YSO associations and CO clouds has already been performed, and details can be found in \citet{2024MNRAS.529.1091Z}, where the radial velocities as well as their uncertainties have been derived from the line profile of the CO emissions. We used several Gaussian components to fit the CO line profile and derived the central velocities and standard deviation as the velocity and its uncertainty. By checking the spatial distribution of the gas and YSOs visually, we chose the best-matching CO component associated with the YSO association. The central velocity and the uncertainty of the best-matching component are taken as the radial velocity and its error.

\section{Deriving 3D velocities} \label{conversion}
\begin{figure*}
\begin{center}
\includegraphics[scale=0.31]{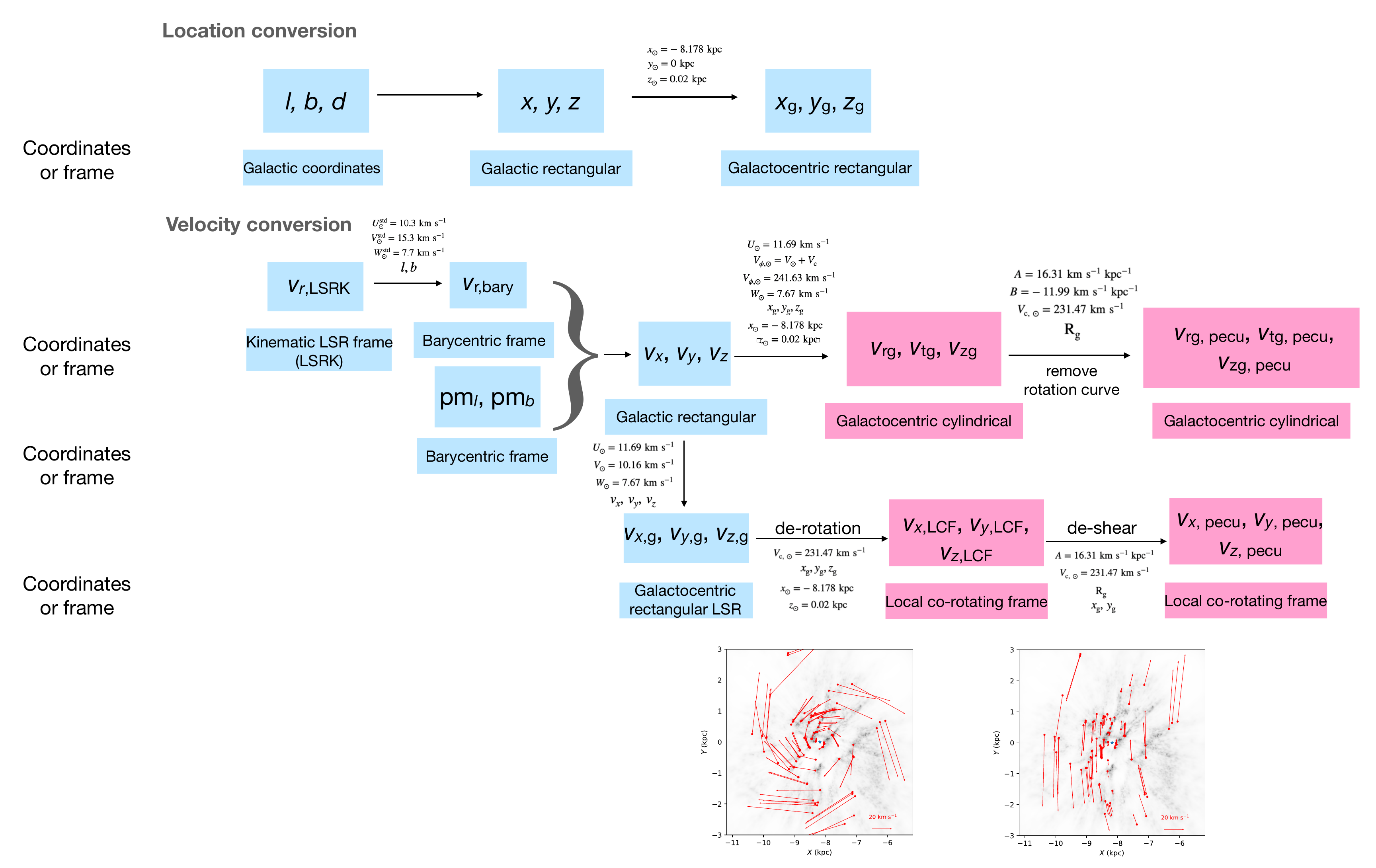}  
\caption{{\bf A diagram of the location and velocity conversion.} We present velocity and location conversions with coordinates or reference frames written below. All the solar location, velocity, and association parameters are noted and given. The inserted subplot below de-rotation shows the rotation field, and the one below de-shear shows the shear field.}
\label{fig1}
\end{center}
\end{figure*}

The aim of this paper is to derive 3D velocities towards our sample YSO-MC complexes by combining CO radial velocity with transverse velocities obtained by the Gaia satellite. Due to historical reasons, past measurements were performed in different frames whose definitions can sometimes be vague. Thus, the first step of our velocity conversion is to convert all measurements to the barycentric frame. The measured velocities are projected into different frames for convenience. These details will be discussed in this section. A list of all the reference frames can be found in the \textcolor{blue}{App. \ref{frames}}, and the conversion process is illustrated in Fig. \ref{fig1}.

% Last three frame are newly defined by us for easy and clear velocity calculation. In Fig.\ref{conversion}, we mark all the reference frame below different velocities and locations.

% (Overview, transformation diagram,{achieve consistency},{convince})

\subsection{Location conversions}
We adopt two sets of locations for the YSO-MC complexes:

\begin{itemize}
    \item $(x, y, z)$ in Galactic rectangular coordinates. In this coordinate, the $x$ axis is pointed from the Sun to the Galactic centre, and the $y$ axis is perpendicular to the $x$ axis and points to the direction that follows the Galactic rotation. The $z$ axis points to the North Galactic Pole.  The Sun is located at (0, 0, 0). The Galactic centre is located at (8.178 kpc, 0, 0) \citep{2019A&A...625L..10G}. $(x, y, z)$ is related to $(l, b, d)$ by $x = d \; {\rm cos}(b)\; {\rm cos}(l),\ y = d\; {\rm cos}(b)\; {\rm sin}(l),\ z = d\;{\rm sin}(b)$ where $l$ and $b$ is Galactic longitude and latitude, and the $d$ is the distance from the Sun. 
    \item (x$_{\rm g}$, y$_{\rm g}$, z$_{\rm g}$) in Galactocentric rectangular coordinates. In this coordinate, the Sun is located at (-8.178 kpc, 0 kpc, 0.02 kpc) \citep{2019MNRAS.482.1417B, 2019A&A...625L..10G} and the Galactic centre is located at (0, 0, 0). $x_{\rm g}$ axis points from the Sun to the Galactic centre, the $y_{\rm g}$ axis points from the Galactic centre to $l = 90^{\circ}$, and the $z_{\rm g}$ axis roughly points to the North Galactic Pole.
    To convert from $(x, y, z)$ to $(x_{\rm g}, y_{\rm g}, z_{\rm g}$), one needs the location of the Sun: ${\rm x}_\odot$ and solar height above the Galactic midplane ${z\rm }_\odot$. We adopt ${\rm x}_\odot = -8.178\ \rm kpc$, and ${\rm z}_\odot = 0.02\ \rm kpc$ \citep{2019MNRAS.482.1417B, 2019A&A...625L..10G}.
    
\end{itemize} 

\begin{figure*}
\begin{center}
\includegraphics[scale=0.26]{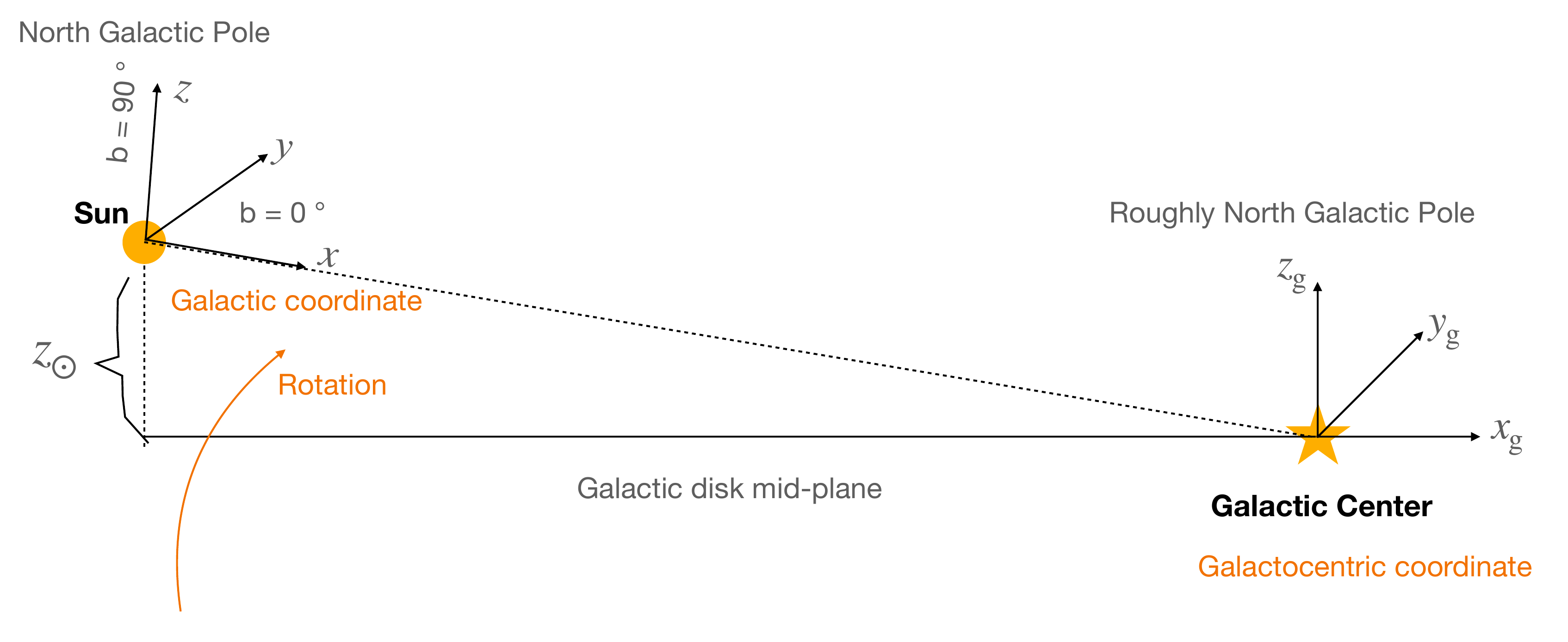}  
\caption{{\bf A comparison between Galactic rectangular and Galactocentric rectangular coordinates.}  The orange star and the orange dot mark the location of the Galactic centre and the Sun, respectively. The orange curve shows the direction of Galactic rotation.}
\label{location frame}
\end{center}
\end{figure*}

These two sets of coordinates are shown in Fig.\ref{location frame}. The angle between the Galactocentric rectangular coordinates and the Galactic rectangular coordinates is ${\rm arctan}(z_{\odot}/x_{\odot}$).

\subsection{Velocity conversions}
The computation of 3D velocities consists of a few key steps:
\begin{enumerate}
    \item Convert radial velocity measurements from the kinematic LSR frame to the Barycentric frame, where the Gaia astrometry measurements are represented.  
    \item Derive 3D velocities in the Barycentric frame.
    \item Convert velocity measurements to other frames.

\end{enumerate}
\subsubsection{Converting CO velocities into the Barycentric frame}
The proper motions and locations of YSO-MC complexes are provided by Gaia DR3 and are measured in the barycentric frame, whose velocity is measured with respect to the centre of gravity of the Solar System.

The radial velocity of each CO cloud is derived from the best-fitting CO emission and has already been converted to the kinematic Local Standard of Rest coordinates (LSRK)\footnote{\url{https://docs.astropy.org/en/stable/api/astropy.coordinates.LSRK.html}}. To derive the 3D velocity, we first convert this measurement back to the barycentric frame. This conversion requires a modification of the Standard Solar Motion in the line of sight. We use the following equation from \cite{2009ApJ...700..137R}:

\begin{small}
\begin{equation}
 v_{\rm r, bary} = v_{\rm r,LSRK} - (U^{\rm Std}_{\odot} \cos l + V^{\rm Std}_{\odot} \sin l) \cos b - W^{\rm Std}_{\odot} \sin b,
\end{equation}
\end{small}

\noindent where $v_{\rm r, bary}$ is the radial velocity in the barycentric frame, $l$ and $b$ are the mean Galactic longitude and latitude of the YSO-MC complex, and ($U^{\rm Std}_{\odot}$, $V^{\rm Std}_{\odot}$, $W^{\rm Std}_{\odot}$) is the standard solar motion towards the Galactic centre, $l = 90^{\circ}$, and the North Galactic Pole, respectively. Following \cite{2009ApJ...700..137R}, we adopt the standard solar motion values: $U^{\rm Std}_{\odot} = 10.3 \, \rm km \, s^{-1}$, $V^{\rm Std}_{\odot} = 15.3 \, \rm km \, s^{-1}$, $W^{\rm Std}_{\odot} = 7.7 \, \rm km \, s^{-1}$, which are consistent with the values used when reducing radio observations \citep{2022PASP..134k4501C}.

\subsubsection{Velocities in the Galactic rectangular coordinate frame}
The next step is to convert velocities to the Galactic rectangular coordinate frame, where the Sun is at the origin $(0, 0, 0)$ and the Galactic centre is at $(8.178, 0, 0)$ kpc. 

We use the \texttt{Galpy} package \citep{2015ApJS..216...29B} to calculate the 3D velocity in the Galactic rectangular frame for each YSO-MC complex. The inputs are the line-of-sight velocity, the proper motions in the $l$ and $b$ directions ($\mu_l$, $\mu_b$) measured in the barycentric frame, as well as the location information, e.g., $l$, $b$, and $d$. The output is the velocity ($v_{\rm x}$, $v_{\rm y}$, $v_{\rm z}$) measured in the Galactic rectangular coordinate frame.

\subsubsection{Velocities in the Galactocentric cylindrical frame and derivation of the peculiar velocity} \label{vrtz}
The Galactocentric frames have their origins at the centre of the Galaxy. To convert our velocities into these frames, we perform a rotation to align the $x$-axis with the mid-plane of the Galactic disc, add a translation to move the origin of our new coordinate system to the centre of the Galaxy, and subtract the solar motion relative to the Galactic centre. Our input data include ($v_x, v_y, v_z$) in the Galactic rectangular frame, ($x_{\rm g}$, $y_{\rm g}$, $z_{\rm g}$) in the Galactocentric rectangular frame, the solar motion $\vec{v}_{\odot}$ relative to the Galactic centre, $x_{\odot}$, and the solar height above the Galactic midplane $z_{\odot}$. We adopt the solar motion ($U_{\odot} = 11.69 \, \rm km \, s^{-1}$, $V_{\odot} + V_{\rm c} = 241.63 \, \rm km \, s^{-1}$, $W_{\odot} = 7.67 \, \rm km \, s^{-1}$) in the Galaxy \citep{2021MNRAS.504..199W}.

The output is ($v_{\rm rg}$, $v_{\rm tg}$, $v_{\rm zg}$) in the Galactocentric cylindrical frame, which represents the circular motion of the YSO-MC complex in the Galaxy. (There is a minus sign in front of $v_{\rm tg}$, and the rotation direction is defined by the right-hand rule, corresponding to the Galaxy's clockwise rotation.)

In the Galactocentric cylindrical frame, we derive the peculiar velocity of each YSO-MC complex by subtracting the rotation motion from the tangential velocity component. The velocity rotation gradient in the solar neighborhood is defined by $-(A+B)$ \citep{1927BAN.....3..275O}, where $A$ and $B$ are Oort constants with values of $16.31 \, \rm km \, s^{-1} \, kpc^{-1}$ and $-11.99 \, \rm km \, s^{-1} \, kpc^{-1}$, respectively. The rotation motion is calculated as $v_{\rm rotation} = v_{\rm c, \odot} - (A+B) \times (r - r_\odot)$, where $r$ is the distance of each YSO-MC complex to the Galactic centre, $r_\odot = 8.178 \, \rm kpc$, and $v_{\rm c, \odot} = 231.47 \, \rm km \, s^{-1}$ \citep{2021MNRAS.504..199W}. The peculiar velocity $\vec{v}_{\rm peculiar}$ in Galactocentric cylindrical coordinates is $(v_{\rm rg}, v_{\rm tg} - v_{\rm rotation}(r_{\rm gal}), v_{\rm zg})$ for a YSO-MC complex at a Galactic distance of $r_{\rm gal}$.

\subsubsection{Velocities in the LSR and the Local Co-rotation Frame (LCF)}
The LSR (Local Standard of Rest) frame is where the mean velocity of stars in the solar neighborhood vanishes. The LSR frame is an inertial frame whose centre rotates with the Galaxy. It is still one of the frames where different velocity measurements are compared.  

We also derive velocities in the Local Co-rotating Frame (LCF, \citealt{2022MNRAS.516L..35L}), which follows the mean motion of stars with an angular velocity of $v_{\rm circ} / r_{\rm gal}$, where $v_{\rm circ}$ is the solar circular velocity at $r_{\rm gal}$, and $r_{\rm gal}$ is the solar distance from the centre of the Milky Way, such that the $x$-axis of the LCF always points to the centre of the Galaxy. This frame is non-inertial, and the Coriolis force must be accounted for. Studying the motion of stars and gas in this frame is convenient as the angles of its axes are locked to the Galactic centre. This frame is similar to those used in, e.g., shearing box simulations of discs.  

\begin{itemize}
\item \textbf{Transform velocities in the Galactic rectangular coordinate frame to the Galactocentric rectangular LSR frame:}

To convert velocities to the LSR frame, we account for the following: First, the Sun moves with respect to the LSR frame with a velocity of ($U_{\odot}$, $V_{\odot}$, $W_{\odot}$). Second, the Sun is located 20 pc above the disc mid-plane. 

To correct for the solar motion, we subtract the solar velocity relative to the LSR frame: ($U_{\odot} = 11.69 \, \rm km \, s^{-1}$, $V_{\odot} = 10.16 \, \rm km \, s^{-1}$, $W_{\odot} = 7.67 \, \rm km \, s^{-1}$) \citep{2021MNRAS.504..199W}. This produces the 3D velocity of the YSO-MC complex in the Galactic rectangular LSR frame.

Next, we perform a rotation to align the $xy$-plane with the Galactic midplane by modifying the solar location ($x_{\odot} = -8.178 \, \rm kpc$, $y_{\odot} = 0 \, \rm kpc$, $z_{\odot} = 0.02 \, \rm kpc$) \citep{2019MNRAS.482.1417B, 2019A&A...625L..10G}. The result is the 3D velocity ($v_{x, \rm g}$, $v_{y, \rm g}$, $v_{z, \rm g}$) for YSO-MC complexes in the Galactocentric rectangular LSR frame.

\item \textbf{Velocities in the Local Co-rotating Frame (LCF)} \label{de-ro}
After deriving the 3D velocity ($v_{x, \rm g}$, $v_{y, \rm g}$, $v_{z, \rm g}$) for YSO-MC complexes in the Galactocentric rectangular LSR frame, we define a new frame---the Local Co-rotating Frame (LCF) \citep{2022MNRAS.516L..35L}, which rotates around the Galactic centre with its $x$-axis fixed toward the centre of the Galaxy. 

Compared to the LCF, the Galactocentric rectangular LSR frame rotates with an angular velocity $\vec{\Omega}_{\rm frame} = (0, 0, -\frac{v_{\rm c, \odot}}{R_{\odot, \rm g}})$, where $v_{\rm c, \odot} = 231.47 \, \rm km \, s^{-1}$ and $R_{\odot, \rm g} = 8.178 \, \rm kpc$ \citep{2019A&A...625L..10G, 2021MNRAS.504..199W}. The rotation speed at a point $(x_{\rm g}, y_{\rm g}, z_{\rm g})$ is related to the distance from the Sun: $\vec{\Delta r} = (x_{\rm g} + 8.178 \, \rm kpc, y_{\rm g}, z_{\rm g} - 0.02 \, \rm kpc)$, and is calculated using: $\vec{v}_{\rm LSR, rotation} = \vec{\Omega}_{\rm frame} \times \vec{\Delta r}$. 

The 3D velocity in the Local Co-rotating Frame (LCF) is: $\vec{v}_{\rm LCF} = \vec{v}_{\rm g, LSR} - \vec{v}_{\rm LSR, rotation}$. The subtracted rotation is shown in the lower-left panel of Fig.~\ref{fig1} and is projected onto the $xy$-plane.
\end{itemize}

\section{Results}

\subsection{YSO-MC Complex Sample with 3D Velocity in Different Reference Frames}
We calculate full 3D velocities of the YSO-MC complexes in different reference frames, as illustrated in Sec. \ref{conversion}. The velocities are shown in Tab. \ref{table1:example}, which includes all velocities and their corresponding errors in different reference frames. Errors are derived using \texttt{Galpy}, with the input errors being the proper motion of YSOs and the radial velocities from CO data. 

In our sample, we highlight some well-known molecular clouds, such as the Orion, $\lambda$ Orion, Perseus, Taurus, Ophiuchus, Lupus, and Chamaeleon molecular clouds. For those strongly associated with gas, we present their 3D velocities in Tab. \ref{table2:clouds}.

\begin{table*}
\begin{tabular}{|c|c|c|c|c|c|c|c|c|c|c|c|}
\hline
YSO-MC ID    & 0       & 1       & 2       & 3       & 4       & 5       & 6       & 7       & 8       & 9    & ...   \\ \hline
$l\ (^\circ)$           & 171.67  & 169.56  & 168.68  & 174.47  & -21.80  & -23.58  & -20.33  & -7.12   & -6.74   & -6.80    & ... \\ \hline
$b (^\circ)$           & -15.46  & -15.70  & -15.88  & -15.00  & 8.99    & 8.56    & 9.35    & 19.48   & 18.91   & 17.01   & ... \\ \hline
$d\ (\rm kpc)$           & 0.13    & 0.13    & 0.13    & 0.14    & 0.16    & 0.16    & 0.16    & 0.14    & 0.14    & 0.14   & ... \\ \hline
pm$_l\ ({\rm mas\ yr^{-1}})$          & 21.85   & 23.19   & 23.98   & 19.80   & -23.40  & -23.58  & -23.23  & -24.13  & -24.14  & -24.10   & ...\\ \hline
pm$_b\ ({\rm mas\ yr^{-1}})$         & -9.69   & -10.83  & -11.19  & -7.35   & -10.55  & -10.55  & -10.58  & -10.12  & -10.72  & -11.54  & ... \\ \hline
$v_{r,\ {\rm bary}}\ ({\rm km\ s^{-1}})$     & 14.91   & 15.64   & 15.41   & 15.65   & 0.16    & -0.53   & -0.34   & -7.82   & -6.57   & -5.10   & ... \\ \hline
$v_{r,\ {\rm err}}\ ({\rm km\ s^{-1}})$       & 1.30    & 1.30    & 1.30    & 1.30    & 1.95    & 1.30    & 2.60    & 1.30    & 0.65    & 0.00     & ...\\ \hline
$x_{\rm g}$ (kpc)          & -8.30   & -8.30   & -8.30   & -8.32   & -8.03   & -8.04   & -8.03   & -8.05   & -8.05   & -8.05   & ... \\ \hline
$y_{\rm g}$ (kpc)         & 0.02    & 0.02    & 0.02    & 0.01    & -0.06   & -0.06   & -0.05   & -0.02   & -0.02   & -0.02   & ... \\ \hline
$z_{\rm g}$ (kpc)         & -0.01   & -0.02   & -0.02   & -0.02   & 0.04    & 0.04    & 0.05    & 0.07    & 0.07    & 0.06  & ...  \\ \hline
$v_x\ ({\rm km\ s^{-1}})$          & -14.60  & -15.64  & -15.58  & -15.05  & -5.22   & -6.43   & -5.18   & -7.07   & -5.77   & -4.51    & ...\\ \hline
$v_y\ ({\rm km\ s^{-1}})$          & -11.73  & -11.90  & -11.94  & -12.03  & -16.79  & -16.33  & -16.74  & -15.44  & -15.44  & -15.46   & ...\\ \hline
$v_z\ ({\rm km\ s^{-1}})$          & -9.84   & -10.77  & -10.85  & -8.86   & -7.78   & -7.84   & -7.92   & -9.01   & -8.86   & -8.77     & ...\\ \hline
$v_{x,\ {\rm err}}\ ({\rm km\ s^{-1}})$       & 1.24    & 1.23    & 1.23    & 1.25    & 1.79    & 1.18    & 2.41    & 1.22    & 0.61    & 0.02     & ...\\ \hline
$v_{y,\ {\rm err}}\ ({\rm km\ s^{-1}})$     & 0.25    & 0.31    & 0.29    & 0.26    & 0.72    & 0.53    & 0.90    & 0.24    & 0.20    & 0.21     & ...\\ \hline
$v_{z,\ {\rm err}}\ ({\rm km\ s^{-1}})$      & 0.36    & 0.37    & 0.36    & 0.35    & 0.31    & 0.21    & 0.43    & 0.44    & 0.23    & 0.12    \\ \hline
$v_{\rm rg}\ ({\rm km\ s^{-1}})$         & 3.45    & 4.61    & 4.59    & 3.75    & -8.08   & -6.98   & -8.02   & -5.06   & -6.34   & -7.60     & ...\\ \hline
$v_{\rm tg}\ ({\rm km\ s^{-1}})$          & -229.90 & -229.72 & -229.67 & -229.60 & -224.79 & -225.25 & -224.84 & -226.18 & -226.17 & -226.15  & ...\\ \hline
$v_{z\rm g}\ ({\rm km\ s^{-1}})$       & -2.13   & -3.06   & -3.14   & -1.16   & -0.10   & -0.15   & -0.23   & -1.33   & -1.17   & -1.09    & ... \\ \hline
$v_{\rm rg,\ err}\ ({\rm km\ s^{-1}})$       & 1.24    & 1.23    & 1.23    & 1.25    & 1.79    & 1.18    & 2.41    & 1.22    & 0.61    & 0.02   & ... \\ \hline
$v_{\rm tg,\ err}\ ({\rm km\ s^{-1}})$      & 0.25    & 0.31    & 0.29    & 0.26    & 0.72    & 0.53    & 0.90    & 0.24    & 0.20    & 0.21    & ... \\ \hline
$v_{z\rm g,\ err}\ ({\rm km\ s^{-1}})$       & 0.36    & 0.37    & 0.36    & 0.35    & 0.31    & 0.21    & 0.43    & 0.44    & 0.23    & 0.12   & ...  \\ \hline
$v_{\rm rg,\ pecu}\ ({\rm km\ s^{-1}})$     & 3.45    & 4.61    & 4.59    & 3.75    & -8.08   & -6.98   & -8.02   & -5.06   & -6.34   & -7.60    & ... \\ \hline
$v_{\rm tg,\ pecu}\ ({\rm km\ s^{-1}})$    & -1.03   & -1.21   & -1.27   & -1.28   & -7.31   & -6.83   & -7.27   & -5.86   & -5.86   & -5.89   & ... \\ \hline
$v_{z\rm g,\ pecu}\ ({\rm km\ s^{-1}})$    & -2.13   & -3.06   & -3.14   & -1.16   & -0.10   & -0.15   & -0.23   & -1.33   & -1.17   & -1.09   & ... \\ \hline
$v_{x,\ \rm g}\ ({\rm km\ s^{-1}})$        & -2.94   & -3.97   & -3.92   & -3.38   & 6.45    & 5.24    & 6.49    & 4.60    & 5.90    & 7.16   & ... \\ \hline
$v_{y,\ \rm g}\ ({\rm km\ s^{-1}})$          & -1.57   & -1.74   & -1.78   & -1.87   & -6.63   & -6.17   & -6.58   & -5.28   & -5.28   & -5.30   & ... \\ \hline
$v_{z,\ \rm g}\ ({\rm km\ s^{-1}})$          & -2.13   & -3.06   & -3.14   & -1.16   & -0.10   & -0.15   & -0.23   & -1.33   & -1.17   & -1.09   & ... \\ \hline
$v_{x,\ \rm g,\ err}\ ({\rm km\ s^{-1}})$      & 1.24    & 1.23    & 1.23    & 1.25    & 1.79    & 1.18    & 2.41    & 1.22    & 0.61    & 0.02   & ... \\ \hline
$v_{y,\ \rm g,\ err}\ ({\rm km\ s^{-1}})$     & 0.25    & 0.31    & 0.29    & 0.26    & 0.72    & 0.53    & 0.90    & 0.23    & 0.20    & 0.21   & ...  \\ \hline
$v_{z,\ \rm g,\ err}\ ({\rm km\ s^{-1}})$   & 0.36    & 0.37    & 0.36    & 0.35    & 0.31    & 0.21    & 0.43    & 0.44    & 0.23    & 0.12   & ... \\ \hline
$v_{x,\ \rm LCF}\ ({\rm km\ s^{-1}})$     & -3.46   & -4.62   & -4.61   & -3.76   & 8.09    & 7.00    & 8.03    & 5.06    & 6.34    & 7.60    & ... \\ \hline
$v_{y,\ \rm LCF}\ ({\rm km\ s^{-1}})$    & -5.14   & -5.29   & -5.26   & -5.76   & -2.53   & -2.15   & -2.42   & -1.54   & -1.56   & -1.56   & ... \\ \hline
$v_{z,\ \rm LCF}\ ({\rm km\ s^{-1}})$     & -2.13   & -3.06   & -3.14   & -1.16   & -0.10   & -0.15   & -0.23   & -1.33   & -1.17   & -1.09   & ... \\ \hline
$v_{x,\ {\rm pecu}}\ ({\rm km\ s^{-1}})$ & -3.45   & -4.61   & -4.60   & -3.75   & 8.13    & 7.04    & 8.07    & 5.07    & 6.35    & 7.61    & ... \\ \hline
$v_{y,\ {\rm pecu}}\ ({\rm km\ s^{-1}})$ & -1.02   & -1.20   & -1.25   & -1.27   & -7.25   & -6.78   & -7.21   & -5.85   & -5.85   & -5.87   & ... \\ \hline
$\sigma_{\rm 2D}\ ({\rm km\ s^{-1}})$    & 1.94    & 1.88    & 1.34    & 2.05    & 1.92    & 2.72    & 0.82    & 1.85    & 1.71    & 1.67   & ...\\ \hline
\end{tabular}
\caption{{\bf 3D velocities for YSO-MC complexes.} The location, proper motion,
radial velocity, and 3D velocities and errors in different coordinates are
presented in the table. On a small portion of the table is published here. The
full version is available online, also at \url{https://doi.org/10.5281/zenodo.16364877}.}
\label{table1:example}
\end{table*}

\begin{table*}
\begin{tabular}{|c|c|c|c|c|c|c|c|}
\hline
cloud name    & Taurus  & Lupus   & Ophiuchus & Chamaeleon & Perseus & Orion   & $\lambda$ Orion \\ \hline
$l\ (^\circ)$            & 171.67  & -21.80  & -7.12    & -56.42     & 158.45  & -151.64 & -165.46         \\ \hline
$b (^\circ)$            & -15.46  & 8.99    & 19.48    & -14.67     & -20.74  & -18.44  & -11.39          \\ \hline
$d\ (\rm kpc)$           & 0.13    & 0.16    & 0.14     & 0.20       & 0.29    & 0.39    & 0.40            \\ \hline
pm$_l\ ({\rm mas\ yr^{-1}})$          & 21.85   & -23.40  & -24.13   & -20.54     & 11.31   & 0.87    & 2.30            \\ \hline
pm$_b\ ({\rm mas\ yr^{-1}})$          & -9.69   & -10.55  & -10.12    & -6.61      & -2.71   & 0.49    & 0.27            \\ \hline
$v_{r,\ {\rm bary}}\ ({\rm km\ s^{-1}})$      & 14.91   & 0.16    & -7.82    & 10.06      & 10.30   & 23.14   & 25.48           \\ \hline
$v_{r,\ {\rm err}}\ ({\rm km\ s^{-1}})$       & 1.30    & 1.95    & 1.30     & 1.30       & 1.30    & 2.60    & 1.95            \\ \hline
$x_{\rm g}$ (kpc)            & -8.30   & -8.03   & -8.05    & -8.07      & -8.43   & -8.51   & -8.56           \\ \hline
$y_{\rm g}$ (kpc)            & 0.02    & -0.06   & -0.02    & -0.16      & 0.10    & -0.18   & -0.10           \\ \hline
$z_{\rm g}$ (kpc)           & -0.01   & 0.04    & 0.07     & -0.03      & -0.08   & -0.10   & -0.06           \\ \hline
$v_x\ ({\rm km\ s^{-1}})$           & -14.60  & -5.22   & -7.07    & -11.43     & -13.48  & -18.80  & -23.18          \\ \hline
$v_y\ ({\rm km\ s^{-1}})$            & -11.73  & -16.79  & -15.44   & -17.40     & -11.51  & -11.99  & -10.53          \\ \hline
$v_z\ ({\rm km\ s^{-1}})$           & -9.84   & -7.78   & -9.01    & -8.51      & -7.16   & -6.45   & -4.52           \\ \hline
$v_{x,\ {\rm err}}\ ({\rm km\ s^{-1}})$        & 1.24    & 1.79    & 1.22     & 0.71       & 1.14    & 2.17    & 1.85            \\ \hline
$v_{y,\ {\rm err}}\ ({\rm km\ s^{-1}})$        & 0.25    & 0.72    & 0.24     & 1.05       & 0.56    & 1.17    & 0.49            \\ \hline
$v_{z,\ {\rm err}}\ ({\rm km\ s^{-1}})$        & 0.36    & 0.31    & 0.44     & 0.33       & 0.48    & 0.83   & 0.39            \\ \hline
$v_{\rm rg}\ ({\rm km\ s^{-1}})$           & 3.45    & -8.08   & -5.06    & -4.64      & 4.54    & 2.35    & 8.83            \\ \hline
$v_{\rm tg}\ ({\rm km\ s^{-1}})$            & -229.90 & -224.79 & -226.18  & -224.18    & -230.09 & -229.74 & -231.22         \\ \hline
$v_{z\rm g}\ ({\rm km\ s^{-1}})$           & -2.13   & -0.10   & -1.33    & -0.81      & 0.55    & 1.27    & 3.21            \\ \hline
$v_{\rm rg,\ err}\ ({\rm km\ s^{-1}})$        & 1.24    & 1.79    &1.22     & 0.71       & 1.14    & 2.17    & 1.85            \\ \hline
$v_{\rm tg,\ err}\ ({\rm km\ s^{-1}})$       & 0.25    & 0.72    & 0.23     & 1.05       & 0.56    & 1.17   & 0.49            \\ \hline
$v_{z\rm g,\ err}\ ({\rm km\ s^{-1}})$       & 0.36    & 0.31    & 0.44     & 0.33       & 0.48    & 0.83    & 0.39            \\ \hline
$v_{\rm rg,\ pecu}\ ({\rm km\ s^{-1}})$      & 3.45    & -8.08   & -5.06    & -4.64      & 4.54    & 2.35    & 8.83            \\ \hline
$v_{\rm tg,\ pecu}\ ({\rm km\ s^{-1}})$      & -1.03   & -7.31   & -5.86    & -7.73      & -0.28   & -0.30   & 1.40            \\ \hline
$v_{z\rm g,\ pecu}\ ({\rm km\ s^{-1}})$      & -2.13   & -0.10   & -1.33    & -0.81      & 0.55    & 1.27    & 3.21            \\ \hline
$v_{x,\ \rm g}\ ({\rm km\ s^{-1}})$          & -2.94   & 6.45    & 4.60     & 0.24       & -1.81   & -7.13   & -11.50          \\ \hline
$v_{y,\ \rm g}\ ({\rm km\ s^{-1}})$          & -1.57   & -6.63   & -5.28    & -7.24      & -1.35   & -1.83   & -0.37           \\ \hline
$v_{z,\ \rm g}\ ({\rm km\ s^{-1}})$          & -2.13   & -0.10   & -1.33    & -0.81      & 0.55    & 1.27    & 3.21            \\ \hline
$v_{x,\ \rm g,\ err}\ ({\rm km\ s^{-1}})$       & 1.24    & 1.79    & 1.22     & 0.71       & 1.14    & 2.17    & 1.85            \\ \hline
$v_{y,\ \rm g,\ err}\ ({\rm km\ s^{-1}})$      & 0.25    & 0.72    & 0.24     & 1.05       & 0.56    & 1.17    & 0.49            \\ \hline
$v_{z,\ \rm g,\ err}\ ({\rm km\ s^{-1}})$      & 0.36    & 0.31    & 0.44     & 0.33       & 0.48    & 0.83    & 0.39            \\ \hline
$v_{x,\ \rm LCF}\ ({\rm km\ s^{-1}})$      & -3.46   & 8.09    & 5.06     & 4.72       & -4.65   & -2.12   & -8.70           \\ \hline
$v_{y,\ \rm LCF}\ ({\rm km\ s^{-1}})$      & -5.14   & -2.53   & -1.54    & -4.26      & -8.53   & -11.10  & -11.15          \\ \hline
$v_{z,\ \rm LCF}\ ({\rm km\ s^{-1}})$      & -2.13   & -0.10   & -1.33    & -0.81      & 0.55    & 1.27    & 3.21            \\ \hline
$v_{x,\ {\rm pecu}}\ ({\rm km\ s^{-1}})$ & -3.45   & 8.13    & 5.07     & 4.79       & -4.55   & -2.35   & -8.84           \\ \hline
$v_{y,\ {\rm pecu}}\ ({\rm km\ s^{-1}})$ & -1.02   & -7.25   & -5.85    & -7.64      & -0.23   & -0.35   & 1.29            \\ \hline
$\sigma_{\rm 2D}\ ({\rm km\ s^{-1}})$     & 1.94   & 1.92    & 1.85    & 0.82       & 2.43    & 3.06   & 3.19           \\ \hline
\end{tabular}
\caption{{\bf 3D velocities of some well-known molecular clouds.} Location, proper motion, radial velocity, and 3D velocities and errors of Taurus, Lupus, Ophiuchus, Chamaeleon, Perseus, Orion, and $\lambda$ Orion at different coordinates are presented in the table.}
\label{table2:clouds}
\end{table*}

\subsection{Comparison with Galactic Rotation Curve}
\begin{figure}
\begin{center}
\includegraphics[scale=0.54]{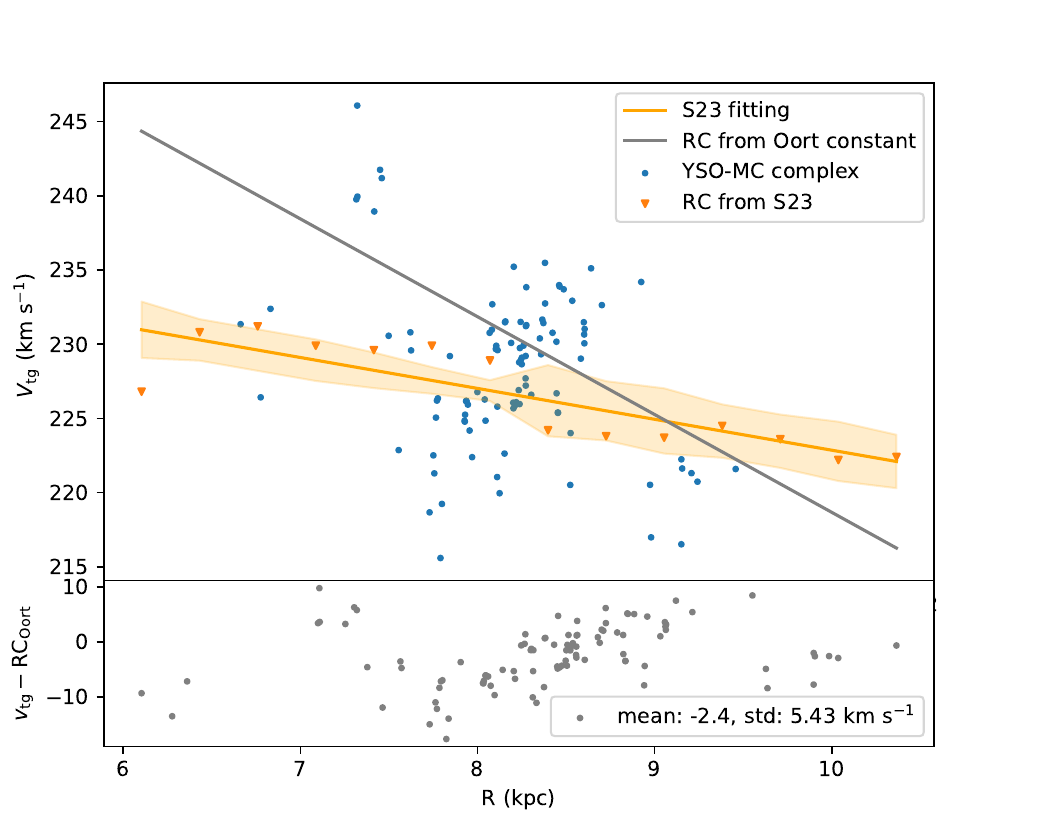}
\caption{{\bf Top panel: Tangential velocity versus Galactic radius for YSO-MC complexes.} The blue dots show the tangential velocity along the Galactic radius from our YSO-MC complexes. The orange triangles are the rotation curve data of the Milky Way derived from Gaia DR3 sources in \citet{2023ApJ...945....3S} (S23), and the orange line is the fitting rotation curve with errors. {\bf Lower panel: Differences between tangential velocities and the rotation curve derived from the Oort constant.} The grey dots are the residual tangential velocity of YSO-MC complexes after the rotation curve is removed. The mean residual velocity is $-$2.4 km s$^{-1}$ and the standard deviation is 5.43 km s$^{-1}$.}
\label{rotation_compare}
\end{center}
\end{figure}

After deriving the velocities of YSO-MC complexes in Galactocentric cylindrical coordinates, we plot their tangential velocity against their Galactic distance in Fig. \ref{rotation_compare}. This figure illustrates how their rotation velocities change with the Galactocentric radius. The calculation of the tangential peculiar velocity is described in Sec. \ref{vrtz}.

We compare our results with the Galactic rotation curve determined from red giant stars and Gaia DR3 data \citep{Eilers2019, Wang2023}, which is combined as Tab. 1 in \citet{2023ApJ...945....3S} (hereafter referred to as S23). The S23 rotation curve is plotted in Fig. \ref{rotation_compare} using orange triangles and fitted with an orange line.

Between $6.5\ {\rm kpc} < R_{\rm g} < 10.5\ {\rm kpc}$, the YSO-MC complexes show a similar decreasing trend in tangential velocity, consistent with the S23 rotation curve. This indicates that these YSO-MC complexes roughly follow the Galactic rotation. We then subtract the assumed rotation velocity at a certain Galactic radius, derived from the Oort constant (represented by the grey line in Fig. \ref{rotation_compare}). The offsets between the rotation velocities of YSO-MC complexes and the rotation curve from Oort constants are small, with a mean value of $-2.4\ {\rm km\ s^{-1}}$.

In Fig. \ref{rotation_compare}, we observe an increasing trend from $R_{\rm g} = 8\ {\rm kpc}$ to $R_{\rm g} = 9\ {\rm kpc}$. We compare the rotation velocity of the YSO-MC complexes with the rotation velocity of 773 Classical Cepheids from \citet{2019ApJ...870L..10M}. This increasing segment aligns with a part of the Cepheids data. In both observations and simulations of Milky Way structures, researchers have identified ridges in the velocity distribution, which are believed to be caused by spiral or bar structures \citep{2018Natur.561..360A, 2018MNRAS.479L.108K, 2019MNRAS.485L.104M}. The increasing part (from 8 to 9 kpc) in our sample coincides with one of the "wiggles" located from 8 to 9 kpc in \citet{2019MNRAS.485L.104M}. Future studies combining these YSO-MC complexes with spiral arm models might help explain the features observed in the figure.

\subsection{Velocity Field}
Our YSO-MC complex sample spans several kiloparsecs on the Galactic plane, enabling us to study the shear velocity field in the solar neighborhood. We begin by considering their barycentric velocities in Galactic rectangular coordinates. We then subtract the frame rotation, Galactic rotation, and shear velocity to derive the peculiar velocity in Galactic rectangular coordinates.

\begin{figure*}
\begin{center}
\includegraphics[scale=0.28]{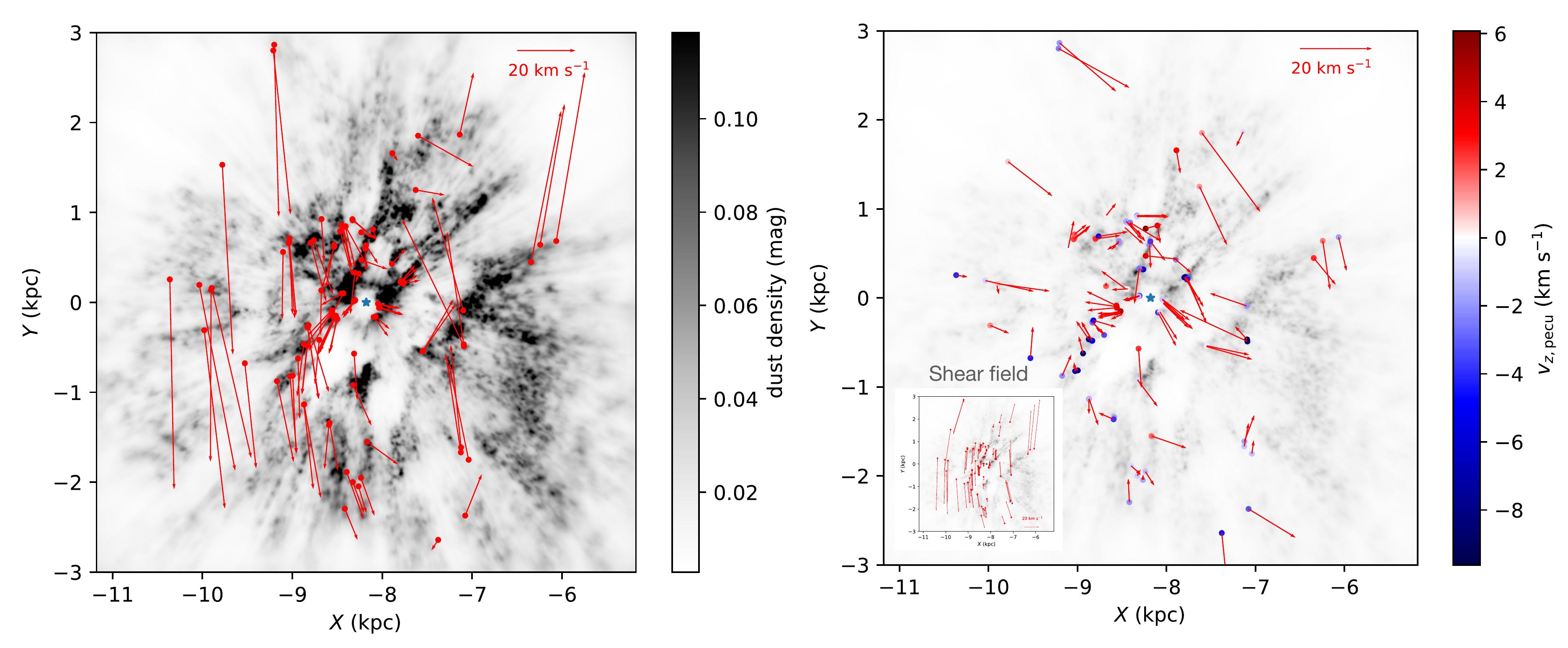}
\caption{{\bf Left panel: $\vec{v}_{\rm LCF}$ of YSO-MC complexes on $XY$ plane.} Red dots refer to the complex location, and the blue star at (-8.17, 0) marks the solar location. The background is the integrated dust map of the solar neighborhood on the $XY$ plane from \citet{2019A&A...625A.135L}. {\bf Right panel: 3D peculiar velocity of YSO-MC complexes on $XY$ plane.} The colors of the scatters represent the peculiar velocities of YSO-MC complexes along the $z$ axis. The inserted map shows the shear field for these YSO-MC complexes. The background is the same dust map as the one in the left panel at a different scale.}
\label{velocity_field}
\end{center}
\end{figure*}

In the top left panel of Fig. \ref{velocity_field}, we plot the velocities of the YSO-MC complexes after removing frame rotation and Galactic rotation. More details about this velocity subtraction can be found in Sec. \ref{de-ro}.

In the solar neighborhood, the shear rate $\kappa$ is estimated using the Oort constant A: $\kappa = 2A$ \citep{1927BAN.....3..275O}. We adopt the constant A of $16.31\ {\rm km\ s^{-1}\ kpc^{-1}}$ from \cite{2021MNRAS.504..199W}. When calculating the shear velocity in the Galactic plane, the solar location is used as the reference. For a given point $(\rm x_g, y_g)$ with Galactocentric distance $\rm r_g = \sqrt{(x_g^2+y_g^2)}$, $\Delta_{\rm r} = r_g - r_{\odot,g}$. The shear velocity $v_{\rm shear}$ is calculated using the formula: ${v}_{\rm shear} = \kappa\Delta_{{\rm r}}$. Considering the angle between the point-Sun direction and the x-axis, $\theta = {\rm arcsin (y_g/r_g)}$, the shear velocity $v_{\rm shear}$ can be decomposed into two components along the x and y directions: $v_{x,\ \rm shear} = {v}_{\rm shear}{\rm sin\theta}$ and $v_{y,\ \rm shear} = {v}_{\rm shear}{\rm cos\theta}$. We plot the shear velocity field for each YSO-MC complex on the $XY$ plane in the bottom right panel of Fig. \ref{fig1} and in the inserted panel in Fig. \ref{velocity_field}.
Its absolute value ranges from 0.85 km s$^{-1}$ to 71.85 km s$^{-1}$ within our sample region. The maximum shear velocity difference along the $Y$ axis is 130.53 km s$^{-1}$. The substantial shear velocity and considerable velocity difference may play an important role in shaping molecular clouds and affecting their evolution and dynamics \citep{1987ApJ...312..626E, 2000ApJ...536..173T, 2014PASJ...66...36M, 2016MNRAS.460.2110F}.

% \begin{figure*}
% \begin{center}
% \includegraphics[scale=0.28]{rotation_curve.pdf}
% \caption{{\bf Left panel: Shear velocity of each YSO-MC complex.} {\bf Right panel: tangential velocities at different Galactocentric radius.}}
% \label{velocity_field}
% \end{center}
% \end{figure*

\subsection{Peculiar Velocity for YSO-MC Complexes}

We calculate the peculiar velocity for each YSO-MC complex in both Galactocentric cylindrical coordinates and Galactic rectangular coordinates. In Galactocentric cylindrical coordinates, we subtract the rotation velocity derived from the Oort constant from the tangential velocity:
$$(v_{\rm rg,\ pecu},\ v_{\rm tg,\ pecu},\ v_{z{\rm g,\ pecu}}) = (v_{\rm rg}, v_{\rm tg}\ -\ v_{\rm rotation}, v_{z\rm {g}}).$$
In Galactic rectangular coordinates, we subtract the frame and Galactic rotation, and the shear field:
$$(v_{x,{\rm pecu}}, v_{y,{\rm pecu}}, v_{z,{\rm pecu}}) = (v_{x,{\rm LCF}}-v_{x,\ {\rm shear}}, v_{y,{\rm LCF}}-v_{y,\ {\rm shear}}, v_{z, {\rm LCF}}).$$
The 3D peculiar velocities of the YSO-MC complexes in both coordinate systems are provided in Tab. \ref{table1:example}.

In the right panel of Fig. \ref{velocity_field}, we present the peculiar velocities of each YSO-MC complex on the $XY$ plane using arrows. The peculiar velocities along the $z$-axis are indicated by colors. The distributions of $v_{x,\ {\rm pecu}}$, $v_{y,\ {\rm pecu}}$ and $v_{z,\ {\rm pecu}}$ are plotted in the left panel of Fig. \ref{percentage}. Along the $x$-axis, the complexes have a mean peculiar velocity of $2.3$ km s$^{-1}$, with a standard deviation of $6.2$ km s$^{-1}$. Towards the $y$-direction, the mean peculiar velocity is $-2.4$ km s$^{-1}$ and the standard deviation of the distribution is $5.7$ km s$^{-1}$. The peculiar velocity along the $z$-axis has a mean value of $-1.4$ km s$^{-1}$ and a standard deviation of $3.1$ km s$^{-1}$. For these YSO-MC complexes, they have a mean 3D peculiar velocity of $8.6$ km s$^{-1}$, which is consistent with the value of $5$ to $10$ km s$^{-1}$ in \citet{Schulz2005} and smaller than the typical value of about $10$ km s$^{-1}$ for maser sources in \citet{2014ApJ...783..130R}. This value can vary slightly across different works due to different solar velocities and is heavily affected by spiral arms \citep{2009ApJ...700..137R, 2013ApJ...769...15X, 2014ApJ...793...72S, 2014A&A...566A..17W, 2018MNRAS.474.2028R}.

\subsection{Anisotropic Velocity}
\begin{figure*}
\begin{center}
\includegraphics[scale=0.67]{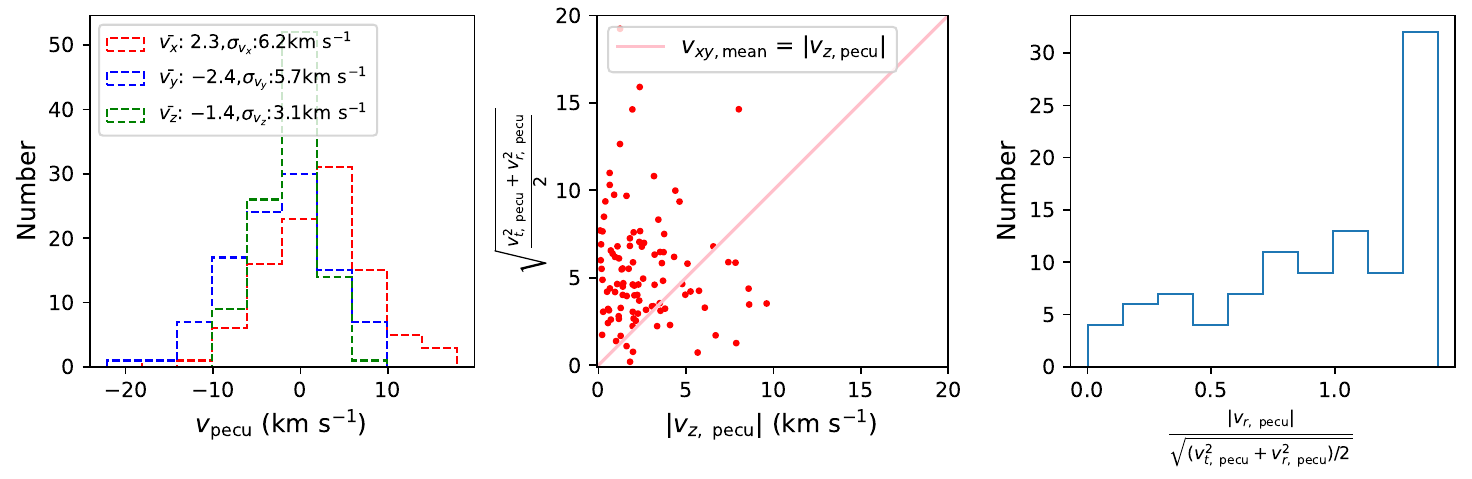} 
\caption{{\bf Left panel: 3D peculiar velocity distribution of YSO-MC.} Red, blue and green histogram refer to the $v_{x,\ {\rm pecu}}$, $v_{y,\ {\rm pecu}}$ and $v_{z,\ {\rm pecu}}$, respectively. The mean velocity and standard deviation in each direction are written in the legend. {\bf Middle panel: $|v_{z,\ {\rm pecu}}|$ vs $\sqrt{\frac{v_{\rm r,\ pecu}^2 + v_{\rm t,\ pecu}^2}{2}}$.} The red dots show the peculiar velocity along the $z$ axis and the predicted peculiar velocity along the $z$ axis under the assumption that energy is distributed equally for the YSO-MC complexes. {\bf Right panel: The distribution of the ratio between absolute radial velocity and the mean velocity on the radial and tangential plane.} The mean velocity on the Galactic plane is calculated from: $\sqrt{(v_{\rm r,\ pecu}^2 + v_{\rm t,\ pecu}^2)/2}$.}
\label{percentage}
\end{center}
\end{figure*}

Assuming that the energy supporting the peculiar velocities is equally distributed along the radial, tangential, and $z$ directions, we predict the peculiar velocity $|v_{z,\ {\rm pred}}| = \sqrt{\frac{v_{\rm rg,\ pecu}^2 + v_{\rm tg,\ pecu}^2}{2}}$, which also serves as the mean velocity on the Galactic plane. We compare the $v_{z,\ {\rm pecu}}$ of the YSO-MC complexes with the predicted $v_{z,\ {\rm pred}}$. The comparison result is plotted in the middle panel of Fig. \ref{percentage}. Compared to the line $y = x$, the $v_{z,\ {\rm pecu}}$ for most of the YSO-MC complexes are lower than the predicted value. This comparison result indicates a lower energy distribution along the $z$-axis than that on the Galactic plane. To study the energy distribution along radial and tangential directions, we examine the ratio between $|v_{r, {\rm pecu}}|$ and the mean peculiar velocity on the Galactic plane. The ratio distribution is plotted in the right panel of Fig. \ref{percentage}. More complexes tend to have a larger peculiar velocity along the radial direction. This observation reveals an uneven energy distribution towards the radial, tangential, and $Z$ directions.

\section{Summary}
Using the 2D velocity and 3D location of YSOs from Gaia DR3 and radial velocity from CO emission of molecular clouds, we calculate 3D velocities for 102 YSO-MC complexes in different reference frames. All the velocities of the YSO-MC complexes and some well-known molecular clouds are provided in the tables. From these velocities, we derive the following results:
\begin{enumerate}
\item From their tangential velocities, these YSO-MC complexes broadly follow the Galactic rotation curve. An increasing substructure at $8-9$ kpc may indicate the ridges found in Cepheid data and simulations.
\item Based on their large spatial coverage, we illustrate the shear velocity field. A maximum $130\ {\rm km}\ {\rm s}^{-1}$ velocity difference is attributed to Galactic shear, which may affect the morphology of molecular clouds and the star formation process.
\item The mean 3D peculiar velocity is $8.6\ {\rm km}\ {\rm s}^{-1}$ for the YSO-MC complexes after subtracting the Galactic rotation and Galactic shear. This value is smaller than, yet consistent with, the peculiar velocity of $10\ {\rm km}\ {\rm s}^{-1}$ from maser measurements.
\item The energy supporting the peculiar velocities is not evenly distributed along different directions. More energy is distributed on the Galactic plane than along the $z$-axis, and more complexes tend to have higher peculiar velocities along the radial direction.
\end{enumerate}

This 3D velocity catalogue of YSO-MC complexes reveals the motion of the clouds and the shear velocity field in the solar neighborhood. When combined with Galactic spiral models and structure simulations, it can help us understand the motion and evolution of the interstellar medium.

\section*{Acknowledgements}
We would like to acknowledge the anonymous referees for their careful reviews and insightful suggestions. JXZ acknowledges support from the China Scholarship Council. GXL acknowledges support from 
NSFC grant 12273032 and 12033005. BQC is supported by the National Key R\&D Program of China No. 2019YFA0405500, National Natural Science Foundation of China 12173034, 11803029, and 11833006, and the science research grants from the China Manned Space Project with NO.\,CMS-CSST-2021-A09, CMS-CSST-2021-A0,8, and CMS-CSST-2021-B03.

This work presents results from the European Space Agency (ESA) space mission Gaia. Gaia data are being processed by the Gaia Data Processing and Analysis Consortium (DPAC). Funding for the DPAC is provided by national institutions, in particular, the institutions participating in the Gaia MultiLateral Agreement (MLA). The Gaia mission website is https://www.cosmos.esa.int/gaia. The Gaia archive website is https://archives.esac.esa.int/gaia. This research has used \texttt{ASTRODENDRO}, a PYTHON package to compute dendrograms from astronomical data (http://www.dendrograms.org/) and \texttt{Galpy}, a PYTHON package for galactic dynamics (https://docs.galpy.org/en/v1.11.0/).

\section*{DATA AVAILABILITY}
The paper used published data from \citet{2016MNRAS.458.3479M} and Gaia DR3 \citep{Gaia2023}. Our Tab. \ref{table1:example} and Tab. \ref{table2:clouds} present the location and 3D velocity information in different reference frames for 102 YSO-MC complexes and some well-known molecular clouds, respectively. Both tables will be available online upon publication.

\bibliographystyle{mnras}
\bibliography{velocity.bib}

\appendix
\section{List of all reference frames}\label{frames}
To clarify, we differentiate between the following frames:

\begin{itemize}
    \item {\bf Barycentric frame:} A frame whose centre is located at the centre
 of gravity of the Solar system.\\
    \item {\bf Local Standard of Rest (LSR) frame:} LSR frame follows the mean motion of stars in the solar neighborhood and centres on the Sun.\\

    \item {\bf Legacy definition of the Local Standard of Rest (LSRK) frame:} This frame is used in radio astronomy. It has a velocity offset from the barycentre of the solar
 system to remove the solar peculiar motion. It can be transformed into
 the barycentric frame by using an old solar velocity in the Galactic
 coordinates: $v_{\rm LSR}=v_{\rm bary}+10.3{\rm cos}(l){\rm cos}(b)+15.3{\rm sin}(l){\rm cos}(b)+7.7{\rm sin}(b)$. More information about LSR frame can be found in \citet{2009ApJ...700..137R} and two websites \footnote{\url{https://www.atnf.csiro.au/people/Tobias.Westmeier/tools_hihelpers.php\#restframes, https://science.nrao.edu/facilities/vla/docs/manuals/obsguide/modes/line}}.
        
    \item {\bf Galactic rectangular coordinate frame:} 
 This is a rectangular
 coordinate system whose origin is in the barycentre of the solar system. The positive
 direction of the x-axis points to the Galactic centre, and the z-axis refers to the
 North Galactic Pole. \\ 
    \item {\bf Galactocentric coordinates frame:} This is a coordinate with
 origin in the Galactic centre. It considers the precise solar location, like
 the distance from the Galactic centre and the height above the Galactic
 midplane, and the solar motion relative to the Galactic centre. \\
    \item {\bf Galactocentric cylindrical coordinates:} This is a cylindrical
 coordinate in the Galactocentric frame, with its origin in the Galactic centre.
 The r-axis refers to the radial direction to the Galactic centre, the t-axis refers to
 the tangential direction, and the z-axis refers to the North Galactic Pole.\\
    \item {\bf Galactic rectangular LSR frame:} This frame centres on the barycentre of
 the solar system with all axes aligned to the Galactic rectangular frame
 and follows the mean motion of stars in the solar neighborhood.\\
    \item {\bf Galactocentric rectangular LSR frame:} This frame is in the Local
 Standard of Rest with all axes aligned to the Galactocentric rectangular
 frame. In this frame, the Sun is located at (-8.178 kpc, 0, 0.02 kpc), and the Galactic centre
 is located at (0, 0, 0).\\
    \item {\bf Local co-rotating frame (LCF):} This frame has its origin in the barycentre
 of the solar system. Except for the circular motion around the Galactic centre, it also contains a rotation around the z-axis with a velocity of an order of $v_{\rm
 circ}/R_{\odot}$, which allows the x-axis to be locked to the Galactic centre. In this
 frame, we can observe how the YSO-MC complex moves.\\
\end{itemize}

More descriptions about coordinates and frames can be found in \url{https://docs.astropy.org/en/stable/coordinates/index.html#overview-of-astropy-coordinates-concepts}, and more conversions can be seen in \url{https://docs.galpy.org/en/stable/reference/util.html}.

\end{document}